# A new interesting source in *Chandra* field: a pulsar wind nebula?

Sudip Bhattacharyya[1]*
[1] *Department of Astronomy and Astrophysics, Tata Institute of Fundamental Research, Mumbai 400005, India*

**ABSTRACT**

We report the detection of a point source CXO J172337.5-373442 in a *Chandra* field with a high significance ($26.7\sigma$), and the discovery ($4\sigma$) of a $48''$ long X-ray tail emanating from the point source. The X-ray spectra of both the point source and the tail are well described with a single absorbed powerlaw, and the tail is harder (power-law index $\Gamma = 0.14^{+0.59}_{-0.68}$) than the point source ($\Gamma = 1.78^{+0.13}_{-0.11}$). From this first detailed spatial, spectral and timing X-ray analysis of CXO J172337.5-373442, and from a plausible optical counterpart found from the archives, we conclude that this source is either a Galactic High-Mass X-ray Binary with an X-ray jet or a Galactic pulsar with its "pulsar wind nebula" seen as the X-ray tail. Although, the currently available data are not enough to distinguish between these two candidates with certainty, a detailed comparison of their known properties with those of CXO J172337.5-373442 favours the latter type. If this identification is correct, then the pulsar should be middle-aged or old, that has escaped from its supernova remnant, and the X-ray tail should originate from the synchrotron emission from either of the following locations: (1) a shocked region, or (2) a jet emanating from the pulsar's magnetosphere.

## 1 INTRODUCTION

The high angular resolution ($0''.5$) and the low background of the Advanced CCD Imaging Spectrometer (ACIS) of the *Chandra* X-ray observatory is ideal to detect faint sources and to minutely study the structures of faint extended sources. ACIS has observed many serendipitous point sources, some of which are attached to relatively faint X-ray tails/jets/lobes. The detection of such faint extended features requires special care during the data analysis. A point source with X-ray tails/jets/lobes can be a protostar, an active galactic nucleus (AGN), an X-ray binary, or a pulsar wind nebula (PWN), and its properties cannot be studied meaningfully without a proper identification. In this Letter, we report the discovery of an X-ray tail emanating from a point source detected with ACIS. From the first detailed X-ray analysis of this source, we conclude that it is either a high-mass X-ray binary (HMXB) system, or a PWN, with the latter type favoured.

## 2 *CHANDRA* DATA ANALYSIS AND RESULTS

The field of the transient low-mass X-ray binary (LMXB) system XTE J1723–376 was observed with the *Chandra* X-ray space mission on 2001 September 4 for 29.7 ks. This observation was made with the ACIS-I mode, "Timed" operating mode, and "Faint" telemetry format. In order to analyse these data, we have used the Chandra Interactive Analysis of Observations (CIAO) software (ver. 4.0; CALDB ver. 3.2.4) starting from the level 1 event file. We have applied the standard grade filtering, removed the high background times, and used the energy range $0.3 - 8.0$ keV and a good exposure time of 29.1 ks for our analysis.

The transient LMXB XTE J1723–376 was in quiescent state, and was not detected with $2\sigma$ threshold. We have used the CIAO tool "celldetect" in order to detect the relatively bright sources in this field. With a threshold of $7\sigma$, "celldetect" has detected only one source. This source has been detected with a very high significance ($\approx 26.7\sigma$; see Table 1; also see later for discussion) at the coordinates (J2000): RA = 17 23 37.532 and Dec = -37 34 41.97. Moreover, smoothing of the image clearly shows that an X-ray tail is attached to this source (Fig. 1; see also Fig. 2). We have not found this X-ray source in the archives, except for the very recently published Brera Multi-scale Wavelet (BMW) *Chandra* source catalogue (Romano et al. 2008). The source 1BMC172337.5-373442 in this catalogue is identical with our source. However, the BMW-*Chandra* source catalogue neither gives any detailed properties of this source, nor mentions the X-ray tail of the source. The high significance of the source and the existence of its tail make it very interesting and worth exploring. Therefore, we study the properties of this source in this Letter. Since we have independently detected this source from the *Chandra* field, we will call it CXO J172337.5-373442 using the *Chandra* naming convention. CXO J172337.5-373442 is located on the ACIS-I3 chip. We have applied the exposure map correction, but found that the effects of non-uniform exposure and non-uniform CCD response in the source region is small. We have also found that the bright portion of this source (that is excluding the tail) is consistent with the point spread function (PSF) of a point source at the $\approx 5'.3$ off-axis location of CXO J172337.5-373442. This strongly suggests that the bright part of CXO J172337.5-373442 is a point source. For our analysis, we have estimated the background from a rectangular region (area $A_{\rm B} = 41054.1$ arcsec$^2$) in the ACIS-I3 chip (see Fig. 1). This region contains no source when the threshold is set to $3\sigma$. We have detected 1037 ($= N_{\rm B}$) counts from this region for the good exposure time, and for the energy range $0.3 - 8.0$ keV. In order to calculate the net count ($N_{\rm N}$) from an area $A_{\rm S}$, we have used the formula $N_{\rm N} = N_{\rm T} - (A_{\rm S}/A_{\rm B})N_{\rm B}$, where $N_{\rm T}$ is the total number of counts detected from that area. Then the error in $N_{\rm N}$ and the signal-to-noise ratio are given by $\Delta N_{\rm N} = \sqrt{N_{\rm N} + (1 + A_{\rm S}/A_{\rm B})(A_{\rm S}/A_{\rm B})N_{\rm B}}$ and $S/N = N_{\rm N}/\Delta N_{\rm N}$ respectively.

* E-mail: sudip@tifr.res.in

We have used the photons detected in a circle of $\approx 6''.8$ radius for the spectral analysis of the point source portion of CXO J172337.5-373442. This circle should contain about 98% of the source counts, as we have found from the instrument PSF. We have detected $714.4 \pm 26.8$ net counts from the 144.3 arcsec$^2$ area of the circle, which implies a signal-to-noise ratio 26.7 (Table 1). We note that the pileup is expected to be very small as the counts per frame is low ($\approx 0.08$). We have used an absorbed powerlaw model (`wabs*powerlaw` of XSPEC) to fit the corresponding energy spectrum. From our fitting (using $\chi^2$-statistic), we have found that the best fit values of the neutral hydrogen column density $N_{\rm H} = 0.37^{+0.10}_{-0.08} \times 10^{22}$ cm$^{-2}$ and the powerlaw photon index $\Gamma = 1.78^{+0.13}_{-0.11}$ (see Table 1). As the corresponding reduced $\chi^2 = 26.1/39$, the model of the observed spectrum does not require any other component (see Fig. 3). Here we note that an absorbed blackbody model (`wabs*bbodyrad` of XSPEC) provides a much worse fit (reduced $\chi^2 = 56.6/39$).

Fig. 1 shows that an X-ray tail starts from the point source portion of CXO J172337.5-373442, and extends roughly towards the north/northwest. The initial part of this faint tail is probably hidden within the PSF of the point source. We have used the photons detected in a polygon (see Fig. 1) of area 436.8 arcsec$^2$ for our analysis. Net counts of $24.0 \pm 5.9$ have been detected from this area, which means that the tail has been detected with a significance of $4.0\sigma$. Fig. 1 indicates that the X-ray tail bends towards northeast near its end. The length of the tail is $\approx 48''$. The tail intensity has a minimum at a location $\approx 21''$ from the point source, a maximum $\approx 10''$ away from the minimum, and after that the tail fades. The average width of the tail is $\approx 12''$. The PSF of the point source portion of CXO J172337.5-373442 cannot explain such size and structure, which strongly suggests that the tail is real. In order to show that the brightness variation and the kink of the tail may be real (and not a result of the smoothing; see Fig. 1), we have displayed the unbinned and unsmoothed *Chandra* image of the CXO J172337.5–373442 region in Fig. 2.

We have fitted the tail spectrum with an absorbed powerlaw model (`wabs*powerlaw` of XSPEC). As the number of counts is small, we have used C-statistic, and fixed the $N_{\rm H}$ to a value of $\approx 0.37 \times 10^{22}$ cm$^{-2}$ (which is the best fit value for the point source). The spectral analysis suggests that the X-ray tail is harder than the point source (Table 1). In order to check it in a model independent way, we have compared the hardness ratios ($h/s$) of the point source and the tail. Here $h$ is the number of net counts in the $2 - 8$ keV energy range and $s$ is that in the $0.3 - 2$ keV range. We have found that $h/s = 0.749 \pm 0.057$ and $1.788 \pm 0.929$ respectively for the point source and the tail. This, like the results of the spectral analysis, indicates that the tail is spectrally harder. From the spectral fitting we have also found that the tail is about 14 times fainter than the point source (see Table 1).

We have done the timing analysis of the point source portion of CXO J172337.5-373442. With a frequency bin of 0.034 milli-Hertz, we have found a marginal feature at the frequency 0.0631 Hz (period $\approx 15.9$ s) with the peak power of 18.95 and the quality factor of 1874. This period is much smaller than the dither periods (1000 s in Y and 707 s in Z directions), and is not an integer multiple of the lowest time bin size (3.24104 s) of the data. Therefore, this timing feature cannot be a result of dithering or the minimum time bin size. The single trial significance of the feature is $3.95\sigma$. However, it is not significant if we consider the number of trials. Nevertheless, we mention it in this Letter, because a known frequency may lead to a significant detection from the future observational data.

## 3 MULTIWAVELENGTH VIEW OF CXO J172337.5–373442

An archival search did not yield any radio, ultraviolet or $\gamma$-ray counterpart of CXO J172337.5–373442. However, we have found a pointlike optical and infrared (IR) source about $0''.15$ away (which we will call "source A"). This is the only optical and IR source in the *Chandra* ACIS error circle of the point source portion of CXO J172337.5–373442. The observed $B$, $V$, $R$, $J$, $H$ and $K$ magnitudes of source A are 16.79, 15.63, 15.31, 12.58, 11.86 and 11.60 respectively.

We will now try to understand the nature of source A from its observed optical colours. First we note that the $B - V$ values of almost all the quasars (for a sample of 788 quasars; see the Fig. 1.6 of Peterson 1997) are less than 1, while $B - V = 1.16$ for source A. Moreover, CXO J172337.5–373442, which is likely a Galactic source (see § 4), is the only possible X-ray counterpart of source A. Therefore source A is not likely a quasar. Hence it is likely a Galactic source, as it is a point source.

In order to check if source A is a star, we have calculated its $E(R - V)/E(B - V)$ ratio assuming that it is a star (considering various stellar spectral types one by one). In order to calculate this ratio we have used the observed $B - V$ and $R - V$ colours of source A, and the known in-trinsic colours $(B-V)_0$ and $(R-V)_0$ of various spectral types of stars. The standard value of $E(R-V)/E(B-V)$ is $-0.78$ (Table 3.21 of Binney & Merrifield 1998). We have not found a value for any stellar spectral type that is close to this standard value. The closest value ($\approx -0.33$) is for O9 supergiants, and the discrepancy is much more for low mass stars. Therefore, source A is plausibly not a star, and at least is very unlikely to be a low mass star.

Since CXO J172337.5–373442 may be a pulsar (see § 4), we will now examine if the optical colour of source A is consistent with that of a pulsar. Unlike the stars, the intrinsic optical colours of the pulsars are poorly explored. Therefore we will compare source A with two well known pulsars: Crab and Vela. The intrinsic colours $(B-V)_0$ of Crab and Vela are 0.08 and 0.185 respectively (Sollerman et al. 2000; Golden et al. 2000; Mignani & Caraveo 2001; Shibanov et al. 2003). If we assume the same intrinsic colours for source A, then we get $E(B-V) = 1.08$ and 0.975, implying the neutral atomic hydrogen column density $N_\mathrm{H} = 0.52 \times 10^{22}$ and $0.47 \times 10^{22}$ respectively (inferred from $N_\mathrm{H} = 0.48 \times 10^{22} E(B-V)$ cm$^{-2}$; Table 3 of Bohlin et al. 1978). These are much less than the Galactic neutral hydrogen column density $1.47 \times 10^{22}$ cm$^{-2}$ in this direction (Dickey & Lockman 1990; NASA's HEASARC $N_\mathrm{H}$ tool), implying that the optical colour of source A is consistent with that of Galactic pulsars.

From the observed optical and IR source density in the CXO J172337.5–373442 region, we find that the probability of the existence of an unrelated optical and IR source within the *Chandra* error circle by chance is about 0.01. This, and the obervational indication that source A is a Galactic source, suggest that this source is the optical and IR counterpart of CXO J172337.5–373442. Assuming this to be true, we have plotted a multiwavelength spectrum of CXO J172337.5–373442 (Fig. 4). However, we note that detailed optical and IR observations will be necessary to confirm this assumption.

## 4 IDENTIFICATION OF CXO J172337.5–373442

The newly detected source CXO J172337.5–373442 is not the transient LMXB XTE J1723–376, because *ASCA* measured the position of the latter (during its outburst with $0'.5$ accuracy; Marshall et al. 1999), which is $\approx 5'$ away from the former. *ASCA* did not detect CXO J172337.5–373442, because (1) the expected $7.762 \times 10^{-3}$ count rate of *ASCA* GIS, and $8.065 \times 10^{-3}$ count rate of *ASCA* SIS were not enough to significantly detect this source for the exposure of $\approx 10$ ks; and (2) the PSF of the bright XTE J1723–376 hindered the detection.

The $4\sigma$ tail is the most striking feature of CXO J172337.5–373442. Therefore, we primarily use this tail in order to identify the source. The tail could be a jet from a protostar. However, we reject this option as no star forming region is found (in optical or IR) at the source location. The tail could also be a jet from an AGN. X-ray jets are common in such systems (Harris & Krawczynski 2006). However, we think that CXO J172337.5–373442 is very unlikely to be an AGN for the following reasons. (1) No host galaxy is seen (in optical or IR) at the source location. (2) The Galactic coordinates of the source are L = 350.250209, B = $-0.824652$, which shows that the source is very close to the Galactic plane and bulge. It should be difficult to detect an AGN in such a direction because of relatively high absorption. (3) Many AGN spectra show a soft excess below 2 keV (above that expected from a simple powerlaw; Mushotzky et al. 1993). But the spectrum of CXO J172337.5–373442 is adequately fitted with a single (absorbed) powerlaw, and does not require any additional soft X-ray component (see § 2 and Fig. 3). (4) More importantly, *Chandra* data analysis gives $N_\mathrm{H} = 0.37^{+0.10}_{-0.08} \times 10^{22}$ cm$^{-2}$ (§ 2), while the total Galactic neutral hydrogen column density in this direction is $1.47 \times 10^{22}$ cm$^{-2}$ (§ 3). This strongly suggests that CXO J172337.5–373442 is a Galactic source. Here we note that to the best of our knowledge the neutral hydrogen data in the source direction is not sufficient to measure the source distance. However, since the source Galactic latitude is low, we assume that the aforesaid $1.47 \times 10^{22}$ cm$^{-2}$ value is for a column length of $10 - 20$ kpc (which is consistent with the data given in Diplas & Savage 1994). With this, and with the assumption that the neutral hydrogen density is uniform for a length scale $> 1$ kpc, the observed $N_\mathrm{H} = 0.37^{+0.10}_{-0.08} \times 10^{22}$ cm$^{-2}$ gives a crude source distance of $2 - 6$ kpc.

CXO J172337.5–373442 could also be an X-ray binary system, because X-ray jets have been detected from such sources (Liu et al. 2006; 2007). However, this source may not be an LMXB, because its X-ray spectrum is relatively hard and plausibly entirely nonthermal (§ 2; Bhattacharya & van den Heuvel 1991). Moreover, if source A is the optical counterpart of CXO J172337.5–373442, then it cannot be an LMXB, because (1) source A is very unlikely to be a low mass star (§ 3); and (2) optical to X-ray luminosity ratio for LMXBs is less than 0.1 (Bhattacharya & van den Heuvel 1991), while this ratio is greater than 1 for CXO J172337.5–373442 (see Fig. 4). CXO J172337.5–373442 can be an HMXB, because the observed optical flux (assuming source A is the optical counterpart) is greater than the X-ray flux. However, to the best of our knowledge, so far only one (out of 114) Galactic HMXB is known to have extended X-ray jets/lobes (SS 433; Liu et al. 2006). This suggests that the probability is low for CXO J172337.5–373442 to be an HMXB. The following points strengthen this tentative conclusion to some extent. (1) SS 433 has two X-ray lobes on the opposite sides of the central source (Migliari et al. 2002), while CXO J172337.5–373442 has an X-ray streak (or, tail) on one side (see Fig. 1). (2) The point source component of SS 433 is harder ($\Gamma = 1.40 \pm 0.04$; Namiki et al. 2003) than the point source portion of CXO J172337.5–373442 (see Table 1). (3) The SS 433 lobes are spectrally softer (powerlaw $\Gamma = 2.1 \pm 0.2$; Migliari et al. 2002) than its point source component, unlike CXO J172337.5–373442 (§ 2). (4) If CXO J172337.5–373442 is a Galactic source (see the previous paragraph), then it is at least about 100 times less X-ray luminous than SS 433. (5) if source A is the optical counterpart of CXO J172337.5–373442, then it may not be an HMXB, because source A is plausibly not a star (§ 3).

We think that the point source portion of CXO J172337.5–373442 is likely a pulsar, and the tail is a PWN. This is because many pulsars have PWNe, that are observed as extended X-ray sources (Kargaltsev & Pavlov 2008). These PWNe have various components, including long streaks or tails (Pavlov et al. 2006; Kargaltsev & Pavlov 2008) like the one observed from CXO J172337.5–373442. Moreover, X-ray streaks are conventionally interpreted as PWNe (Muno et al. 2008). In addition, the following points show that the properties of CXO J172337.5–373442 are consistent with those of pulsars and PWNe. (1) The powerlaw index of the X-ray spectrum of the point source part of CXO J172337.5–373442 is consistent with that of pulsars (see Table 1; and Kargaltsev & Pavlov 2008). (2) The powerlaw index of the X-ray spectrum of our source tail is consistent with that of PWNe (e.g., Bhattacharyya S. et al., in preparation). (3) The X-ray point source luminosity to the X-ray tail luminosity ratio ($\approx 14$) of CXO J172337.5–373442 is consistent with that of PWNe, which can be in the range of $\sim 0.1 - 45$ (Kargaltsev & Pavlov 2008; Bhattacharyya S. et al., in preparation). (4) For a source distance of 5 kpc, the X-ray luminosity of the point source portion of CXO J172337.5–373442 is $\sim 1.5 \times 10^{33}$ ergs s$^{-1}$. Therefore, for a reasonable source distance both the point source luminosity and the tail luminosity are very consistent with those observed from pulsars and PWNe (see the Table 2 of Kargaltsev & Pavlov 2008). (5) If the source A is the optical counterpart of CXO J172337.5–373442, then this source can be a pulsar for the following reasons. (a) The optical colour of source A is consistent with that of Galactic pulsars (§ 3). (b) The $N_{\rm H}$ values ($0.47 \times 10^{22}$ cm$^{-2}$ and $0.52 \times 10^{22}$ cm$^{-2}$; § 3) inferred from the pulsar identification of source A are well within the 68.3% and 90% (respectively) confidence limits of the value inferred from the *Chandra* data.

We note that, although CXO J172337.5–373442 can be a pulsar, the identification of source A and CXO J172337.5–373442 as the same pulsar is inconsistent with the fact that, unlike our case, the optical flux of a pulsar is normally less than its X-ray flux. This can be seen from Fig. 4 of Thompson et al. (1999). Therefore, although from the same figure we note that the optical flux may be greater than the X-ray flux for some pulsars (as indicated for PSR B1706-44), CXO J172337.5–373442 cannot be identified as a pulsar with certainty using the currently available data.

## 5  DISCUSSION

In this Letter, we report the detection of a point X-ray source with high significance, and the discovery of a 4$\sigma$ X-ray tail emanating from the point source. This source can be a Galactic HMXB or a Galactic pulsar. From the discussion in § 4, we conclude that the latter identification is favoured. Therefore, we will now briefly discuss some of the properties of this source assuming it to be a pulsar. We notice that the point source (pulsar) X-ray energy spectrum is likely entirely nonthermal (powerlaw; see § 2), which implies that the X-ray emission of the pulsar primarily originates in its magnetosphere (Kargaltsev et al. 2005). This means either the thermal emission from the neutron star's surface is largely obscured (possibly by the plasma in the magnetosphere), or the stellar surface temperature is relatively low. Since no supernova remnant (SNR) is seen near CXO J172337.5–373442, the pulsar has probably escaped from its SNR, implying that it is likely middle-aged or old. As the pulsar is moving through the ISM, its speed may be supersonic as the sound speed in ISM is lower than that in SNR. In such a case, the X-ray taillike PWN may originate from synchrotron emission in a shocked region behind the pulsar (Pavlov et al. 2006). Alternatively, such a tail may be a result of synchrotron process in a jet emanating from the pulsar's magnetosphere along the spin axis (Benford 1984). The observed clumpiness and the bending of the tail (Fig. 1) may be caused by the sausage instability and the kink instability respectively. Finally, we note that if source A and CXO J172337.5–373442 are the same pulsar, then it is a unique pulsar with optical luminosity much higher than the X-ray luminosity. On the other hand, if CXO J172337.5–373442 is an HMXB, then, to the best of our knowledge, this will be the second Galactic HMXB known to have a visible X-ray jet. In order to resolve this, further optical and X-ray observations of the source is essential.

## ACKNOWLEDGMENTS

We thank K. P. Singh and Bhaswati Mookerjea for useful discussion, and an anonymous referee for constructive suggestions.

This paper has been typeset from a TeX/ LaTeX file prepared by the author.

**Table 1.** Properties (with $1\sigma$ error) of various portions of CXO J172337.5–373442.

| Source | Area (arcsec$^2$) | Net count[a] | S/N[b] | Photon index $\Gamma$[c] | Unabsorbed flux[d] ($10^{-13}$ ergs cm$^{-2}$ s$^{-1}$) |
|---|---|---|---|---|---|
| Point source | 144.3 | $714.4 \pm 26.8$ | 26.7 | $1.78^{+0.13}_{-0.11}$ | 4.29 |
| Tail | 436.8 | $24.0 \pm 5.9$ | 4.0 | $0.14^{+0.59}_{-0.68}$ | 0.30 |

[a] Net count (for $0.3 - 8.0$ keV) in the given area.
[b] Signal-to-noise ratio.
[c] For spectral fitting with the XSPEC model `wabs*powerlaw`.
[d] For $0.5 - 10$ keV.

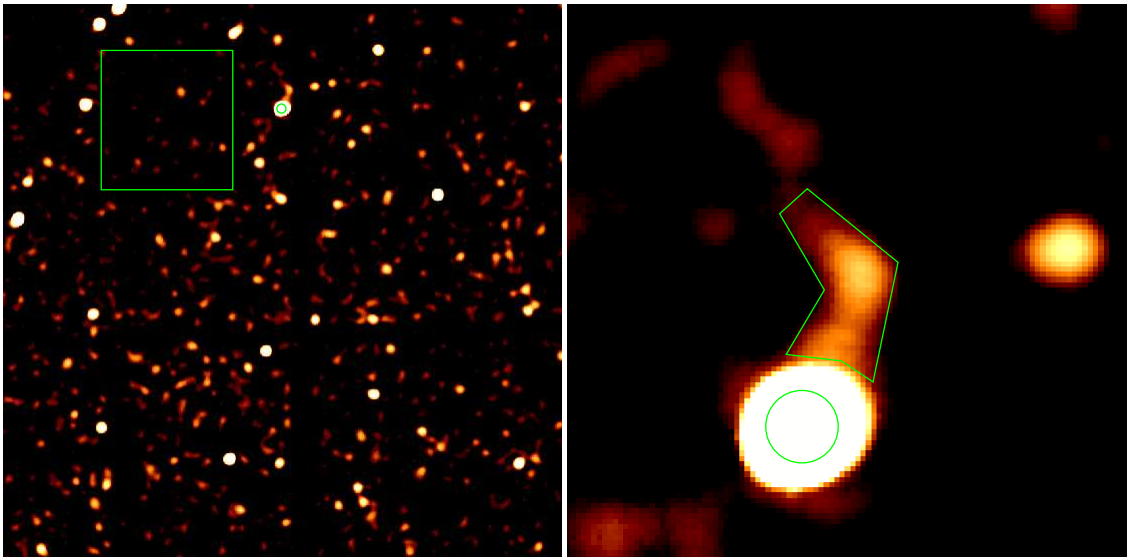

**Figure 1.** Binned (by a factor of 2) and smoothed images. *Left panel*: A $13'.9 \times 13'.8$ *Chandra* field showing the point source portion of CXO J172337.5–373442 with a green circle ($\approx 6''.8$ radius). The background was estimated from the portion depicted with the green rectangle. *Right panel*: Magnified view ($1'.7 \times 1'.7$) of CXO J172337.5–373442. The green polygon encloses the tail, and the green circle ($\approx 6''.8$ radius) shows the point source portion.

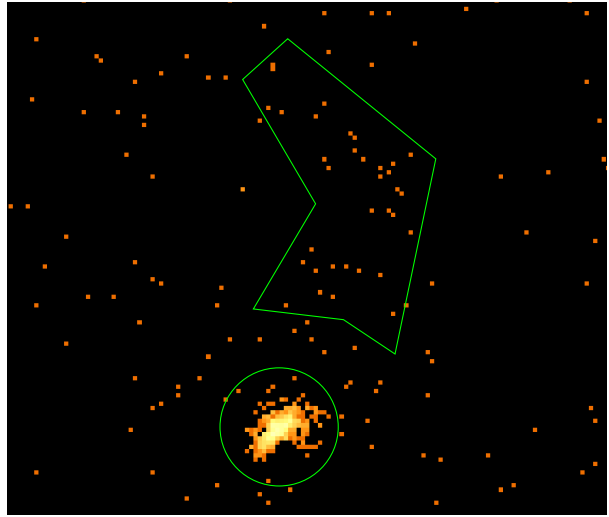

**Figure 2.** An unbinned and unsmoothed $59'' \times 69''$ *Chandra* image of CXO J172337.5–373442 region. The green circle and polygon are same as the ones shown in the right panel of Fig. 1. Note that the shape and size of the central source are consistent with those of the PSF of a point source at this $\approx 5'.3$ off-axis location. This figure suggests that the X-ray brightness variation and the kink of the tail (seen in Fig. 1) are real, and may not be a result of the smoothing.

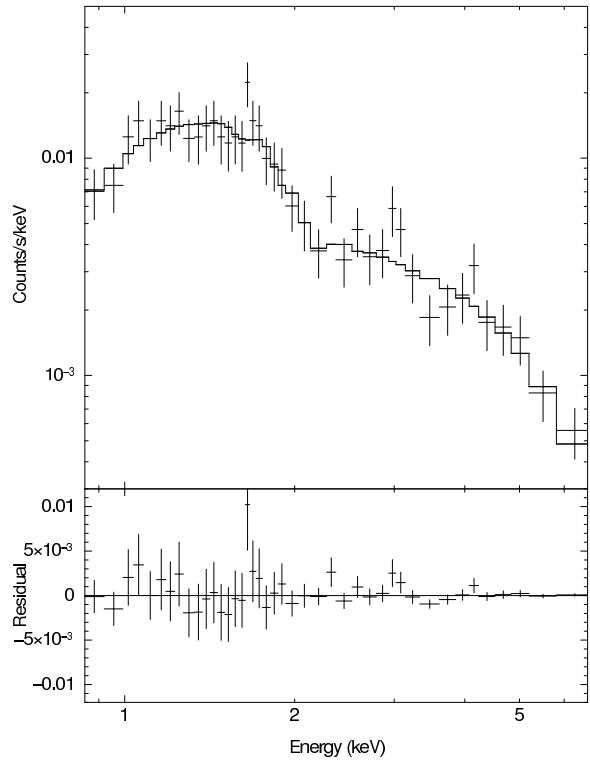

**Figure 3.** *Chandra* ACIS energy spectrum from the point source portion of CXO J172337.5–373442. The data are fitted with the XSPEC model `wabs*powerlaw`. The upper panel shows the data points and the model (solid histogram), and the lower panel shows the residual.

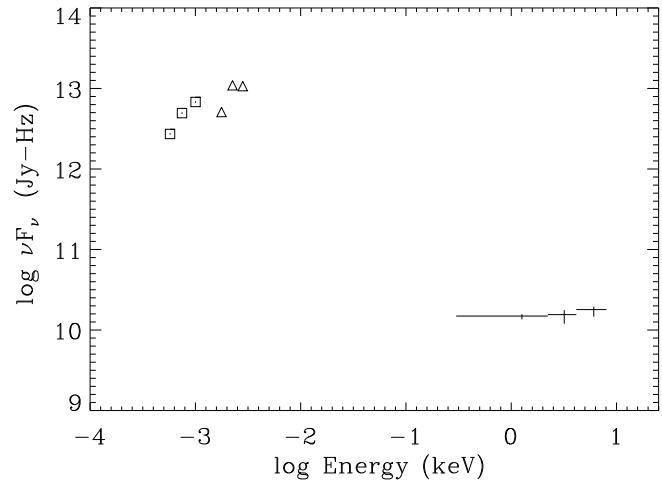

**Figure 4.** A multiwavelength energy spectrum of the point source portion of CXO J172337.5–373442. From the left side, three types of points (three points for each type) are for infrared, optical and X-ray energies. Infrared (J, H and K bands) and optical (B, V and R bands) points are from 2MASS and NOMAD catalogues respectively, and the X-ray points (including the horizontal lines showing the energy ranges) are from the current *Chandra* data analysis. The fluxes of this figure are corrected for absorption using $N_{\rm H} = 0.37 \times 10^{22}$ cm$^{-2}$ and a corresponding $E(B-V) = 0.76$ (see text).